\documentclass[12pt,preprint]{emulateapj}
\usepackage{rotating}
\usepackage{graphicx}
\usepackage{longtable}
\usepackage{ulem}
\usepackage{float}

\newcommand{\be}{\begin{equation}}
\newcommand{\ee}{\end{equation}}

\newcommand{\msun}{{M}_{\sun}}

\slugcomment{Accepted by ApJ Letter on \today}

\shorttitle{Fundamental plane for radiatively efficient black-hole sources} \shortauthors{Dong et al. }

\begin{document}

\title{A new fundamental plane for radiatively efficient black-hole sources}

\author{Ai-Jun Dong\altaffilmark{1}, Qingwen Wu\altaffilmark{1,$\bigstar$} and Xiao-Feng Cao\altaffilmark{1}}

\altaffiltext{1}{School of Physics, Huazhong University of Science and Technology,
 Wuhan 430074, China}
\altaffiltext{2}{$^{\bigstar}$Corresponding author, email: qwwu@hust.edu.cn}

\begin{abstract}
    In recent years, it was found that there are several low/hard state of X-ray binaries (XRBs) follow an `outliers' track of radio--X-ray correlation ($L_{\rm R}\propto L_{\rm X}^{b}$ and $b\sim1.4$), which is much steeper than the former universal track with $b\sim0.6$. In this work, we compile a sample of bright radio-quiet active galactic nuclei (AGNs) and find that their hard X-ray photon indices and Eddington ratios are positively correlated, which is similar to that of `outliers' of XRBs, where both bright AGNs and `outliers' of XRBs have bolometric Eddington ratios $\gtrsim1\%L_{\rm Edd}$ ($L_{\rm Edd}$ is Eddington luminosity). The Eddington-scaled radio--X-ray correlation of these AGNs is also similar to that of `outliers' of XRBs, which has a form of $L_{\rm 5 GHz}/L_{\rm Edd}\propto (L_{\rm 2-10 keV}/L_{\rm Edd})^{c}$ with $c\simeq1.59$ and 1.53 for AGNs and XRBs respectively. Both the positively correlated X-ray spectral evolution and the steeper radio--X-ray correlation can be regulated by a radiatively efficient accretion flow (e.g., disk-corona). Based on these similarities, we further present a new fundamental plane for `outliers' of XRBs and bright AGNs in black-hole (BH) mass, radio and X-ray luminosity space: $\log L_{\rm R}=1.59^{+0.28}_{-0.22} \log L_{\rm X}- 0.22^{+0.19}_{-0.20}\log M_{\rm BH}-28.97^{+0.45}_{-0.45}$ with a scatter of $\sigma_{\rm R}=0.51\rm dex$. This fundamental plane is suitable for radiatively efficient BH sources, while the former plane proposed by Merloni et al. and Falcke et al. may be most suitable for radiatively inefficient sources.

\end{abstract}


\section{Introduction}
    Stellar-mass X-ray binaries (XRBs) and active galactic nuclei (AGNs) seem to have a similar central engine consisting of a central black hole (BH), an accretion disk, and a possible relativistic jet. XRBs spend most of their time in a faint quiescent state, and may undergo sudden and bright several-month-long outbursts with typical recurrence periods of many years. Several spectral states have been identified based on their spectral and timing properties during one outburst \citep[e.g.,][]{MR06}. At the beginning and end of the outburst, XRBs are normally observed in low/hard (LH) state, of which the emission is dominated by a power-law component extending to $\sim$ 100 keV with a photon index of $1.5\lesssim\Gamma\lesssim2$. However, XRBs will stay in high/soft (HS) state at high luminosities, where the X-ray spectrum is characterized by a strong thermal black-body emission and a weak power-law component with $\Gamma\gtrsim2$. Some XRBs also show a very high state (VHS) with $\Gamma\gtrsim2.4$, which differs from the HS state in that the power-law component, rather than the disk component, is dominant (\citealt{MR06,zs13} for recent reviews and references therein). Comparing the different states of XRBs, there are many types of AGNs. The unification of different types of AGNs is generally accepted, where the main driving parameters are orientation of AGNs with respect to the line of sight and radio loudness. The radio loudness is normally defined as ratio of the monochromatic flux density at 5~GHz and optical $B$ band at 4400~$\rm \AA$ ($R_{\rm o}=F_{\rm 5\ GHz}/F_{\rm B}$), where $R_{\rm o}=10$ is usually taken as the division between radio quiet (RQ) and radio loud (RL) AGNs \citep[particularly in quasar studies, e.g.,][]{ke94}. \citet{falc04} suggested that different AGN classes can be identified with corresponding to XRB states based on the power unification scheme, where low luminosity AGNs (LLAGNs) are analog to LH state of XRBs, RQ quasars are analog to HS state of XRBs.

   The different BH activities as observed in XRBs and AGNs are believed to be triggered by different accretion processes. Both soft X-ray bumps observed in HS state of XRBs and optical/UV bumps observed in quasars can be naturally interpreted by multi-temperature blackbody emission from a cold, optically thick, geometrically thin standard accretion disk (SSD; \citealt{SS73}). The prevalent accretion model for LH state of XRBs and LLAGNs is a hot, optically thin, geometrically thick radiatively inefficient accretion flow (RIAF) that has been developed for BH accreting at low mass accretion rate (e.g., \citealt{nara94}; and see \citealt{nara08} and \citealt{ho08} for recent reviews). \cite{wugu08} found that hard X-ray photon indices are anti-correlated to Eddington ratios for XRBs when the Eddington ratios are less than a critical value ($\sim$ 1\%), and, however, they become positively correlated when the Eddington ratios are higher than this critical value, where the anti- and positive correlations also exist in LLAGNs \citep[e.g.,][]{gc09,con09} and bright AGNs \citep[e.g.,][]{wang04,sh08} respectively. This phenomena are consistent with predictions of RIAF model and disk-corona model respectively very well (e.g., \citealt{cao09,qiao13}).

   The quasi-simultaneous radio and X-ray fluxes of LH-state XRBs roughly follow a `universal' non-linear correlation of $F_{\rm R}\propto F_{\rm X}^{b}$ ($b\sim0.5-0.7$) as initially found in GX 339-4 and V404 Cyg \citep[][]{corb03,gall03}. By taking into account the mass of BH, the relation was extended to AGNs, which is called ``fundamental plane'' of BH activity \citep{merl03,falc04,kord06,plot12}. The plane can be described by \citep[e.g.,][]{merl03}
   \be
   \log L_{\rm R}=0.60^{+0.11}_{-0.11}\log L_{\rm X}+0.78^{+0.11}_{-0.09}\log M_{\rm BH}+7.33^{+4.05}_{-4.07},
   \ee
   where $L_{\rm R}$ is 5 GHz nuclear radio luminosity in unit of $\rm erg\ s^{-1}$, $L_{\rm X}$ is 2-10 keV nuclear X-ray luminosity in unit of $\rm erg\ s^{-1}$, and $M_{\rm BH}$ is the BH mass in unit of $\msun$. The radio spectrum of XRBs is usually flat or even inverted, which is often taken as evidence for presence of jets \citep[][]{fen01}. The X-ray emission of XRBs is normally thought to originate in accretion flows \citep[e.g., disk/corona and/or RIAF,][]{yc05} or even jet base \citep[e.g.,][]{ma05}. Regardless of its physical explanation, the non-linear `universal' scaling implies that the same mechanism governing the accretion and ejection processes from BHs holds over approximately nine orders of magnitude in mass.

   However, \citet{xc07} found that the radio--X-ray correlation in XRBs might not be as universal as previously thought. In the following years, more and more XRBs were found to lie well outside the scatter of the former universal radio--X-ray correlation (e.g., H1743$-$322, \citealt{jonk10,cori11}; Swift 1753.5$-$0127, \citealt{cado07,sole10}; XTE J1752$-$223, \citealt{ratt12}). These outliers roughly form a different `outliers' track, which follow a steeper radio--X-ray correlation with an index of $b\sim 1.4$ as initially found in H1743$-$322 \citep{cori11}. Some of these sources (e.g., H1743$-$322, XTE J1752$-$223, MAXI J1659$-$152) jump to the standard universal correlation when they fade towards quiescence \citep{jonk10,cori11,ratt12}. \citet{cw14} found that the radio--X-ray correlation is tightly correlated to the X-ray spectral evolution, where the data points with an anti-correlation of $\Gamma-F_{\rm X}$ will follow the `universal' or `transition' track, while the data points will stay in the `outliers' track if they follow a positive $\Gamma-F_{\rm X}$ correlation. It should be noted that the radio--X-ray correlation of GX 339-4 also become steeper ($b=1.14\pm0.27$) for data points with a positive correlation of $\Gamma-F_{\rm X}$, even though it was regarded as the source fully stay in the universal track in former works.

   Using the refined sample, \citet{kord06} found that the sub-Eddington objects (e.g., LH state of XRBs and LLAGNs) follow the fundamental plane most tightly, and proposed that the fundamental plane of \citet{merl03} and \citet{falc04} should be most suitable for radiatively inefficient BH sources. In this work, we aim to explore the fundamental plane for a sample of radiatively efficient BH sources (e.g., `outliers' of XRBs and bright AGNs),where bolometric Eddington ratio of these sources normally larger than 1\% (e.g., $L_{\rm bol}/L_{\rm Edd}\gtrsim1\%$). The sample is presented in section 2. In section 3, we show several similarities for these two types of BH sources, and then present a new fundamental plane for them. The results are discussed in section 4. Throughout this work, we assume the following cosmology for AGNs: $H_{0}=70\ \rm km\ s^{-1} Mpc^{-1}$, $\Omega_{0}=0.27$ and $\Omega_{\Lambda}=0.73$.

 \section{Sample}

   We consider a sample comprised of radiatively efficient XRBs and AGNs. Only black-hole sources with Eddington ratios larger than a critical ratio (e.g., $L_{\rm bol}/L_{\rm Edd}\gtrsim1\%$) are selected because they may accrete through RIAFs if their Eddington ratios $L_{\rm bol}/L_{\rm Edd}\lesssim1\%$ \citep[e.g.,][for a recent review]{nara08}, where $L_{\rm bol}$ is bolometric luminosity and $L_{\rm Edd}$ is Eddington luminosity.

   For XRBs, we select sources in bright hard state with multiple, quasi-simultaneous (i.e., within a day) radio and X-ray observations. We use the X-ray spectral evolution to isolate the low-hard and bright-hard state XRBs, and select the data points with a positive $\Gamma-F_{\rm X}$ correlation, where these data points have $L_{\rm bol}/L_{\rm Edd}\gtrsim1\%$ and follow the `outliers' track \citep{cw14}. The radio flux, X-ray flux and X-ray photon index of three XRBs (GX 339-4, H 1743-322, and Swift J1753.5-0127, 52 groups of data) are selected from \citet[][and references therein]{cw14}, where the X-ray data are analyzed from $RXTE$ and radio data are obtained by several arrays (e.g., Very Large Array, Australia Telescope Compact Array etc.). The quasi-simultaneous radio and X-ray observations of GRS 1915+105 during the LH state (82 groups of data) are also selected because it's $L_{\rm 3-9keV}/L_{\rm Edd}\gtrsim10\%$ \citep[][]{ru10}, which also roughly follow the `outliers' track \citep[][]{corb13}. The accretion process in all selected data points should be radiatively efficient based on their Eddington ratios and/or positive X-ray spectral evolution. In this work, we use 2-10 keV X-ray luminosity and 5 GHz radio luminosity in exploring their radio--X-ray correlation, where the hard X-ray emission mainly originate from the power-law component due to the disk component is not evident in LH state of XRBs. The radio emission observed in different waveband is extrapolated to 5 GHz assuming a typical radio spectral index of $\alpha=-0.12$ \citep[$F_{\nu}\propto\nu^{-\alpha}$,e.g.,][]{corb13}, where the radio spectrum is normally flat for these LH-state outliers \citep[$|\alpha|\lesssim0.3$, e.g.,][]{cori11,corb13}. The black hole mass $M_{\rm BH}$=12.3, 13.3, 14, 12$\msun$  \citep[][]{zu08,ru13}, and the distances 8, 8, 11, 8 kpc \citep[][and references therein]{corb13} are adopted for GX 339-4, H 1743-322, GRS 1915+105 and Swift J1753.5-0127 respectively.

   For AGNs, we select RQ type I AGNs from CAIXA catalog, of which the Eddington ratio $L_{\rm bol}/L_{\rm Edd}\gtrsim1\%$. The BH masses and radio flux densities are also available \citep[][and references therein]{bi09}, where the radio fluxes are selected from literatures and several sources with radio loudness $R_{\rm o}\gtrsim10$ are excluded. The BH masses of these type I AGNs are calculated from the virial product of the velocity widths of broad H$\beta$ emission line and size of the broad-line region estimated from the empirical size-luminosity relation. The 2--10 keV X-ray luminosities and X-ray photon indices are selected from \citet{zz10}, which are derived from targeted observations from {\it XMM-Newton}. The radio emission from these RQ AGNs mainly originate from compact nuclear region and the contribution from star formation in the host galaxies normally can be neglected \citep[e.g.,][]{zu12}. In total, we have 64 bright RQ AGNs (see Table 1 for more details).

\begin{table*}[t]
\centering
\begin{minipage}{180mm}
\footnotesize
  \centerline{\bf Table 1. The data of RQ AGNs}
\tabcolsep 1.500mm
\begin{tabular}{lcccccc||lcccccc}\hline\hline
\tablecolumns{16}
\tabletypesize{\fontsize{0.4in}}
 Name  &$z$ &  $\Gamma$& $M_{\rm{BH}}$ & $L_{\rm{2-10keV}}$ & $L_{\rm{5 GHz}}$
&$\frac{L_{\rm bol}}{L_{\rm{Edd}}}$  &Name  & $z$ &
$\Gamma$&$M_{\rm{BH}}$ & $L_{\rm{2-10keV}}$ & $L_{\rm{5 GHz}}$
&$\frac{L_{\rm bol}}{L_{\rm{Edd}}}$\\
& & &$M_{\odot}$&ergs/s&ergs/s& & & & &$M_{\odot}$&ergs/s&ergs/s \\
\hline

1H 0419-577    &0.104  & 1.27 &8.58&44.33& 39.83&-0.60 &
Ark 120        &0.032  & 2.02 &8.27&43.95& 38.56& -0.99  \\
Ark 374        &0.063  & 1.94 &7.86&43.49& 38.67& ...&
Ark 564        &0.025  & 2.52 &6.27&43.50& 38.79&  0.07  \\
ESO 323-G77    &0.015  & 1.76 &7.39&42.67& 38.50& ...&
Fairall 9      &0.047  & 1.73 &7.91&43.97& 39.11& -1.72 \\
HE 1029-1401   &0.086  & 1.91 &9.08&44.31& 39.63& -0.88 &
HE 1143-1810   &0.033  & 1.82 &7.01&43.82& 38.61& ... \\
IC 4329A       &0.016  & 1.80 &6.77&43.96& 38.84& -0.83 &
IRAS 1334+2438 &0.108  & 2.05 &8.62&43.81& 40.01& -0.58\\
LZw 1          &0.059  & 2.33 &7.26&43.85& 39.09& 0.12 &
MC-5-23-16     &0.009  & 1.90 &7.85&43.02& 37.68& ... \\
MCG-6-30-15    &0.008  & 1.95 &6.19&42.90& 36.82& -0.81 &
MR2251-178     &0.064  & 1.54 &9.03&44.46& 39.17& ...\\
Mrk 1044       &0.017  & 2.20 &6.50&42.55& 37.52& 0.02 &
Mrk 110        &0.035  & 1.79 &6.82&43.92& 38.16& -0.36 \\
Mrk 1383       &0.087  & 1.99 &8.63&44.10& 38.89& -1.07 &
Mrk 1513       &0.063  & 1.61 &7.58&43.51& 39.00& ... \\
Mrk 205        &0.071  & 1.75 &8.68&43.95& 38.78& -0.57 &
Mrk 279        &0.031  & 1.86 &7.62&43.50& 38.93& -0.81 \\
Mrk 290        &0.030  & 1.59 &7.65&43.25& 38.34& ...&
Mrk 335        &0.026  & 2.28 &7.15&43.27& 38.36& 0.05 \\
Mrk 359        &0.017  & 1.88 &6.24&42.50& 37.69& -0.60 &
Mrk 493        &0.031  & 2.23 &6.17&43.22& 38.05& 0.21 \\
Mrk 509        &0.034  & 1.65 &7.86&44.68& 38.83& -1.10&
Mrk 586        &0.155  & 2.39 &7.55&44.05& 39.81& 0.53 \\
Mrk 590        &0.026  & 1.66 &7.20&42.86& 38.50& -0.16 &
Mrk 766        &0.013  & 2.21 &6.28&43.16& 38.05& -0.26 \\
Mrk 841        &0.036  & 1.95 &7.88&43.89& 38.18& -0.36 &
Mrk 876        &0.129  & 1.84 &7.55&44.19& 39.81& -1.25 \\
NGC 2992       &0.008  & 1.53 &7.72&42.97& 38.83& ...&
NGC 3516       &0.015  & 1.80 &7.36&42.39& 38.58& -1.89 \\
NGC 3783       &0.010  & 1.60 &6.94&43.03& 38.22& -1.36 &
NGC 4051       &0.002  & 2.01 &6.13&41.39& 37.26& -1.50 \\
NGC 4151       &0.003  & 1.65 &7.44&42.22& 38.18& -1.39 &
NGC 4593       &0.009  & 1.69 &6.91&43.07& 37.26& -0.79 \\
NGC 5506       &0.006  & 1.99 &7.46&42.83& 38.90& -0.38 &
NGC 5548       &0.017  & 1.68 &8.03&43.39& 38.70& -1.63 \\
NGC 7314       &0.005  & 2.19 &6.70&42.28& 36.71& ...&
NGC 7469       &0.016  & 1.75 &6.84&43.17& 39.26& -0.49 \\
PDS 456        &0.184  & 2.36 &8.91&44.77& 40.57& -0.11 &
PG 0052+251    &0.155  & 1.83 &8.41&44.61& 39.50& -0.83 \\
PG 0804+761    &0.100  & 1.96 &8.24&44.46& 39.40& -1.07 &
PG 0844+349    &0.064  & 2.16 &8.68&43.74& 38.17& -0.77 \\
PG 0947+396    &0.206  & 1.94 &8.68&44.37& 39.25& -0.93 &
PG 0953+414    &0.234  & 2.12 &8.24&44.73& 40.17& -0.05 \\
PG 1048+342    &0.167  & 1.90 &8.37&44.04& 37.57& -0.85 &
PG 1114+445    &0.144  & 1.46 &8.59&44.16& 38.73& -0.86 \\
PG 1115+407    &0.155  & 2.44 &7.67&43.93& 38.98& ...&
PG 1202+281    &0.165  & 1.75 &8.61&44.43& 38.56& -0.45 \\
PG 1211+143    &0.081  & 1.79 &7.49&43.70& 38.90& -0.56 &
PG 1216+069    &0.331  & 1.67 &9.20&44.72& 40.84& -1.50 \\
PG 1244+026    &0.048  & 2.62 &6.52&43.15& 38.43& 0.12 &
PG 1307+085    &0.155  & 1.52 &7.90&44.08& 39.10& -1.18 \\
PG 1322+659    &0.168  & 2.22 &8.28&44.02& 38.88& -0.24&
PG 1352+183    &0.152  & 1.88 &8.42&44.13& 38.96& -0.30 \\
PG 1402+261    &0.164  & 2.34 &7.94&44.15& 39.56& 0.26 &
PG 1415+451    &0.114  & 2.01 &8.01&43.60& 38.82& -0.77 \\
PG 1416-129    &0.129  & 1.54 &9.05&43.88& 39.93& -1.45 &
PG 1427+480    &0.221  & 1.93 &8.09&44.20& 38.14& -0.41 \\
PG 1440+356    &0.079  & 2.50 &7.47&43.76& 38.88& 0.09 &
PG 1448+273    &0.065  & 2.37 &6.97&43.29& 38.70& 0.43 \\
PG 1626+554    &0.133  & 2.05 &8.54&44.16& 38.66& -0.66 &
RE J1034+396   &0.042  & 2.58 &6.81&42.57& 39.29& 0.16 \\

\hline
\end{tabular}
\end{minipage}

\begin{minipage}{170mm}
 Note: The hard X-ray photon index and 2-10 keV X-ray luminosity are selected from \citet{zz10}, while the BH mass, radio luminosity and Eddington ratio are selected from  \citet{bi09}.
\end{minipage}

\end{table*}

   \section{Results}

  In Figure 1, we present the relation between hard X-ray photon index, $\Gamma$, and Eddington ratio, $L_{\rm bol}/L_{\rm Edd}$ for both AGNs and XRBs. The bolometric luminosity of AGNs is estimated from X-ray luminosity using a luminosity-dependent bolometric correction \citep[see][]{bi09}. The power-law luminosity of $L_{\rm 0.1-200 keV}$ is regarded as the bolometric luminosity for XRBs since that the disk component is not important in LH state of XRBs, where $L_{\rm 0.1-200 keV}$ was extrapolated from the X-ray flux and spectral index that reported in \citet{cw14}. It can be clearly found that both XRBs and AGNs in our sample have $L_{\rm bol}/L_{\rm Edd}\gtrsim$1\%. More importantly, the hard X-ray photon index is positively correlated with the Eddington ratio for AGN sample, which is similar to that of a single XRB, even each XRB seems to follow its own X-ray spectral evolution, where the dashed line is the best fit for AGNs while the solid lines are the best fits for each XRB (see Figure 1).


 \begin{figure}[t]
 \includegraphics[width=80mm]{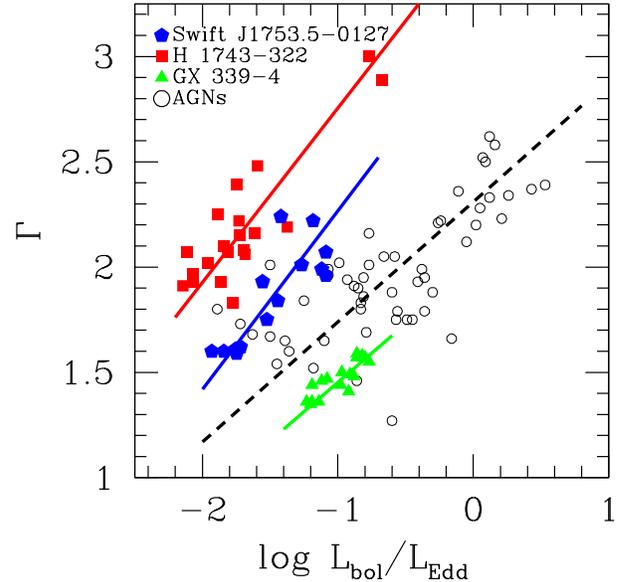}
 \caption{The relation between $\Gamma$ and $L_{\rm bol}/L_{\rm Edd}$ for XRBs and AGNs. Dashed line is best fit for AGN sample while solid lines are the best fits for each XRB.\label{fig1}}

  \end{figure}

\begin{figure}[ht]
\includegraphics[width=80mm]{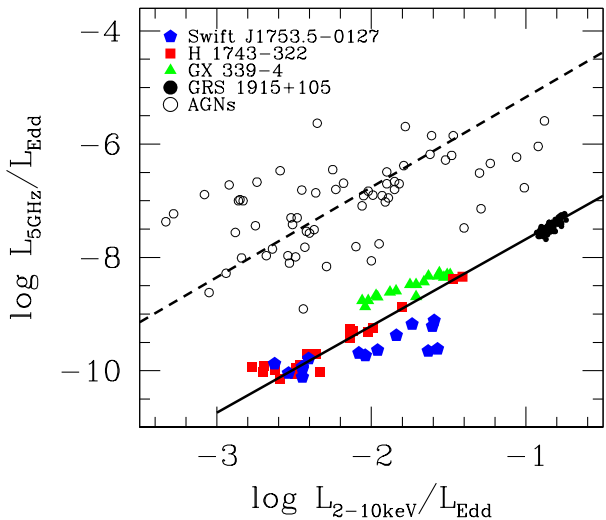}
 \caption{The Eddington-scaled radio--X-ray correlations for AGNs and XRBs, where the dashed and solid lines are the best fits for these two subsamples respectively. \label{fig2}}
\end{figure}

In Figure 2, we plot the relation between Eddington-scaled X-ray luminosity, $L_{\rm 2-10 keV}/L_{\rm Edd}$, and radio luminosity, $L_{\rm 5 GHz}/L_{\rm Edd}$, for AGNs and XRBs respectively, and find that both subsamples show positive correlations. We fit the correlation with a straight line considering the uncertainties in both coordinates. The best fits are
 \be
   \log \left(\frac{L_{\rm 5 GHz}}{L_{\rm Edd}}\right) =1.59\pm0.29 \log \left(\frac{L_{\rm 2-10 keV}}{L_{\rm Edd}}\right) - 3.58\pm0.67,
   \ee
  \be
   \log  \left(\frac{L_{\rm 5 GHz}}{L_{\rm Edd}}\right) = 1.53\pm0.05 \log  \left( \frac{L_{\rm 2-10 keV}}{L_{\rm Edd}}\right) - 6.15\pm0.07,
   \ee
  for the subsamples of AGNs and XRBs respectively, where the typical observational uncertainties $\sigma_{L_{\rm R}}$=0.2 dex \citep[e.g.,][]{hp01}, $\sigma_{L_{\rm X}}$=0.3 dex \citep[e.g.,][]{st05}, and $\sigma_{\rm M}$=0.4 dex \citep[e.g.,][]{vp06} for AGNs and the typical variations (within one day) $\sigma_{L_{\rm R}}$=0.1 dex, $\sigma_{L_{\rm X}}$=0.15 dex\citep[e.g.,][]{cori11,corb13}, and typical uncertainty of BH mass $\sigma_{\rm M}$=0.15 dex \citep[e.g.,][]{zs13} for XRBs are considered. We find the correlation slope of the `outliers' of XRBs is similar to that of RQ AGNs.

  Based on the similarities of hard X-ray spectral evolution and radio--X-ray correlations for these XRBs and AGNs, we further explore the relation between $L_{\rm R}$, $L_{\rm X}$ and $M_{\rm BH}$ for these BH sources with the form, $\log L_{\rm R}=\xi_{\rm X} \log L_{\rm X}+\xi_{\rm M}\log M_{\rm BH}+c_{0}$, as that of \citet{merl03}, where $L_{\rm R}$ is 5 GHz radio luminosity, $\log L_{\rm X}$ is 2-10 keV X-ray luminosity.  To find the multi-parameter relation, we adopt a similar approach as that of \citet{merl03} and minimize the following statistic,
 \be
  \chi^{2}=\sum\limits_{i} \frac{(y_i-c_0-\xi_{\rm X}X_i-\xi_{\rm M}M_i)^2}{\sigma^2_{\rm R}+\xi_{\rm X}^2\sigma^2_{\rm X}+\xi_{\rm M}^2\sigma^2_{\rm M}},
  \ee
  where $y_i$ is measurement of radio luminosities ($\log L_{\rm 5GHz}$), $X_i=\log L_{\rm 2-10 keV}$, $M_i=\log M_{\rm BH}$ and $c_0$ is a constant. Instead of assuming the isotropic uncertainties with $\sigma_{L_{\rm R}}=\sigma_{L_{\rm X}}=\sigma_{\rm M}$ as in \citet{merl03}, we adopt the above typical observational uncertainties/variations for AGNs and XRBs. The best fit for the whole sample of AGNs and XRBs is
  \be
  \log L_{\rm R}=1.59^{+0.28}_{-0.22} \log L_{\rm X}- 0.22^{+0.19}_{-0.20}\log M_{\rm BH}-28.97^{+0.45}_{-0.45},
  \ee
  with $\chi^{2}/\rm d.o.f.=1.32$ and a scatter of $\sigma_{\rm R}=0.51$ dex. To make both samples have a similar distribution of Eddington ratios, we exclude the AGNs with $L_{\rm bol}/L_{\rm Edd}\gtrsim 0.3$ and find that the fitting result is $\log L_{\rm R}=1.56^{+0.35}_{-0.25} \log L_{\rm X}- 0.20^{+0.27}_{-0.29}\log M_{\rm BH}-27.86^{+0.42}_{-0.42}$ with $\chi^{2}/\rm d.o.f.=1.19$ and $\sigma_{\rm R}=0.42$ dex, which is roughly similar to the case with whole AGN sample (equation 5).

\begin{figure}[ht]
\includegraphics[width=80mm]{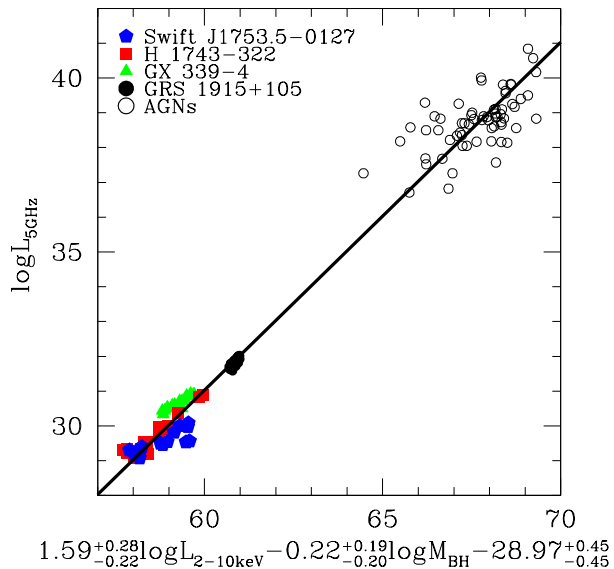}
\caption{The fundamental plane of black hole activity for radiatively efficient XRBs and AGNs. The solid line is best fit for the whole sample. \label{fig3}}
\end{figure}

  \section{Discussion}
  In spite of a vast difference in mass scales, XRBs and AGNs are thought to be powered by an essentially scale-invariant central engine of accretion-jet. The quantitative comparison of XRBs and AGNs has made a major step forwards in last ten years with the discovery of a fundamental plane for BH activity \citep[e.g.,][]{merl03,falc04}.

  However, it has not established definitely how the phenomenology of AGNs can be compared with the different states of XRBs. By the refined sample of AGNs and XRBs, \citet{kord06} proposed that the sub-Eddington AGNs (or LLAGNs) in a state equivalent to the LH state of XRBs follow the fundamental plane most tightly with the intrinsic scatter is comparable to measurement error. Recently, several LH state of XRBs are found to follow the `outliers' track of radio--X-ray correlation with a much steeper slope of $\sim1.4$, which clearly differs from that of the universal one (e.g., 0.5-0.7). \citet{cori11} proposed that accretion processes in these `outliers' may be radiatively efficient, which is supported by their Eddington ratios and the positive X-ray spectral evolution \citep[][]{cw14}. We compile a sample of RQ AGNs with $L_{\rm bol}/L_{\rm Edd}\gtrsim 1\%$ and find that their hard X-ray photon indices are positively correlated to their Eddington ratios, which is similar to that of outliers of XRBs (see Figure 1). We note that each XRB seems to follow individual $\Gamma-L_{\rm bol}/L_{\rm Edd}$ relations, which is also presented in a singe bright AGN \citep[e.g.,][]{so09}. Actually, the positive correlation still statistically exists in XRB sample if we also consider the XRBs in HS state and very high state \citep[][]{wugu08}, which is consistent with the bright AGN sample as a whole. Furthermore, the correlation slope of the Eddington-scaled radio--X-ray correlation is 1.59 and 1.53 for these bright AGNs and `outliers' of XRBs respectively, which are also similar, where the Eddington-scaled radio and X-ray luminosities are less affected by the possible term of BH mass. Similar to the individual X-ray spectral evolution, there are substructures in radio--X-ray correlation of XRBs, where the correlation slope range from $\sim1.1-1.7$ for different outliers \citep[see Figure 2 or][for detailed anlyses]{cw14}, which may be caused by the different properties of accretion flow (e.g., electron temperature, magnetic field etc.). \citet{gall12} re-explored the radio--X-ray correlation for the `outliers' as a whole, and found that the correlation slope is $\sim0.98$, where the shallower correlation may be caused by the data points in the `transition' track are also included in their fitting. For a given X-ray luminosity, the `outliers' show a radio luminosity fainter than that expected from the universal correlation, and, therefore, these `outliers' may be indeed similar to RQ AGNs.

  Based on the similarities of `outliers' of XRBs and bright RQ AGNs, we present a fundamental plane for these radiatively efficient BH sources in BH mass, radio and X-ray luminosity space. The correlation slope $\xi_{\rm X}=1.59$ is much larger than that found in former works \citep[e.g., $\xi_{\rm X}\sim0.6$,][]{merl03,falc04,kord06}. Excluding the AGNs with $L_{\rm bol}/L_{\rm Edd}\gtrsim 0.3$, we find the fundamental plane is roughly unchanged, where AGN sample and XRB sample have more or less similar distributions of Eddington ratios and X-ray photon indices. Our new fundamental plane (equation 5) can be applied to radiatively efficient BH sources while the former fundamental plane \citep[][]{merl03,falc04} may be most suitable for sub-Eddington BH sources. It should be noted that the sample of \citealt{merl03} contained LLAGNs as well as bright quasars. However, \citet{kord06} re-analyzed the samples of \citet{merl03} and \citet{falc04}, and proposed that the fundamental plane for subsample containing only LLAGNs and LH state of XRBs is most tight with a scatter of $\sigma_{\rm R}\lesssim0.2$ dex, which is much smaller than $\sigma_{\rm R}\simeq0.88$ dex for the whole sample with quasars in \citet{merl03}.

   The radiatively inefficient RIAF is expected to produce X-ray emission with $L_{\rm X}\propto \dot{M}^{q}$ ($q\sim 2.0$, e.g., \citealt{merl03,yc05}), which can roughly explain the universal radio--X-ray correlation ($L_{\rm R}\propto L_{\rm X}^{0.5-0.7}$) if considering the scaling between jet luminosity and jet power, $L_{\rm R}\propto Q_{\rm jet}^{1.4}$ (e.g., \citealt{hein03}) and $Q_{\rm jet}\propto \dot{M}$ (e.g., \citealt{falc95}). From simple physical assumptions, the radio--X-ray correlation will become steeper if accretion flow is radiative more efficient than that of RIAFs. The Eddington ratios of the XRBs and AGNs in our sample are normally larger than 1\% (see Figure 1), and, therefore, their accretion process should be radiatively efficient since that the RIAFs only exists in BH sources with $L_{\rm bol}/L_{\rm Edd}\lesssim1\%$. The positive correlation of hard X-ray photon index and Eddington ratio of both `outliers' and bright AGNs is consistent with the prediction of disc-corona model \citep[e.g.,][]{cao09,qiao13}, which suggests that the radiatively efficient accretion disk may exist in these BH sources. In disk-corona model, a fraction $f_c$ of the accretion power is dissipated in corona through magnetic reconnection, and eventually radiates as X-ray emission. The X-ray luminosity of disk-corona model can be written as $L_{\rm X}=f_c\dot{M}c^2$, where $f_c$ is constant when gas pressure is dominant in the cold disk while $f_c$ will decrease as increasing of the accretion rate when radiation pressure becomes dominant \citep[e.g.,][]{mf02}. Therefore, we expect $q\lesssim1.0$ in the disk-corona model. Assuming the jet launching and radiation behave identically in both tracks of XRBs (e.g., $L_{\rm R}\propto Q_{\rm jet}^{1.4}\propto \dot{M}^{1.4}$), we expect a radio--X-ray correlation of the form $L_{\rm R}\propto L_{X}^{\xi}$ and $\xi\gtrsim1.4$ if the jet is launched from the radiatively efficient disk-corona system, which can help to explain the steeper correlation slopes as find in our new fundamental plane. It should be noted that the jet coupled with other radiatively efficient accretion flows \citep[e.g., luminous hot accretion flow, LHAF,][]{xy12} can also explain the observed steeper radio--X-ray correlation. It is still unclear whether the spectral evolution of LHAF is consistent with the observations or not, which is beyond this work.  We note that a flat radio spectrum from compact radio core is crucial for deriving the relation $L_{\rm R}\propto Q_{\rm jet}^{1.4}$ or explaining above correlations. However, the origin of radio emission in RQ AGNs is not entirely clear and it is unknown whether their physical mechanism is similar to that of `outliers' or not. Most of the RQ AGNs are radio compact and occasionally accompanied by linear features plausibly associated with plasma outflows, where more than half of total radio emission originate from the compact radio core \citep[e.g.,][]{ku98}. Normally, the spectrum of radio core is flat in bright RQ AGNs (e.g., NGC 5548, \citealt{wr00} and NGC 4051, \citealt{jon11}), and which may come from the small-scale jet as that of XRBs in LH state.

 \section*{Acknowledgments}
   We thank Xinwu Cao and members of HUST astrophysics group for many useful discussions and comments. This work is supported by the NSFC (grants 11103003, 11133005 and 11303010) and New Century Excellent Talents in University (NCET-13-0238).

\end{document}